\documentclass{aa}
\usepackage{graphicx}
\usepackage{txfonts}
\usepackage{natbib}
\bibpunct{(}{)}{;}{a}{}{,}

\newcommand{\kk}{{\bf k}}
\newcommand{\xx}{{\bf x}}
\newcommand{\yy}{{\bf y}}
\newcommand{\R}{{\bf R}}
\newcommand{\N}{{\bf N}}
\newcommand{\1}{{\bf 1}}

\begin{document}
\title{Detection of cosmic filaments using the Candy model} 
\author{Radu S. Stoica
	\inst{1}
\and Vicent J. Mart\'{\i}nez
\inst{2}
\and Jorge Mateu
\inst{1}
\and Enn Saar
\inst{3}}
	\institute{Departament de Matem\`atiques, Universitat Jaume I, 
	Campus Riu Sec, 12071 Castell\'o, Spain.\\
	\email{stoica@guest.uji.es, mateu@mat.uji.es}
	\and Observatori Astron\`omic de la Universitat de Val\`encia, 
	Apartat de correus 22085, 46071 Val\`encia, Spain.\\
\email{martinez@uv.es}
	\and Tartu Observatoorium, T\~oravere, 61602 Estonia.\\
	\email{saar@aai.ee} }

\offprints{V. J.  Mart\'{\i}nez }

   \date{Received ...; accepted ...}

\abstract{
We propose to apply a marked point process to automatically
delineate filaments of the large-scale structure in redshift
catalogues. We illustrate the feasibility of the idea on an example
of simulated catalogues, describe the procedure, and characterize
the results. We find the distribution of the length of the
filaments, and suggest how to use this approach to obtain other
statistical characteristics of filamentary networks.
}
\maketitle
\section{Introduction}

The large-scale structure of the Universe is studied by creating
galaxy maps -- positions of thousands (a few years ago) and millions
(nowadays) of galaxies in space. The angular positions of galaxies
are relatively easy to measure, but their distances can be estimated
only by measuring their recession velocities. The latter task is
difficult, especially for faint distant objects, and thus really
detailed maps of galaxies have started to appear only lately. An
additional caveat is that the recession velocities contain a
contribution from the dynamical velocity of a galaxy, so the estimated
distances define the so-called 
'redshift space'.

The dominant feature of these maps, as of all other galaxy maps of
the large-scale structure of the universe, is the network of
filaments of different sizes and contrast, along with relatively
empty voids between the filaments. The filamentary network
contains different scales, where smaller-scale filaments are also
less prominent. The gradual disappearance of structures with
increasing distance is due to the fact that the apparent
luminosity of a galaxy is the fainter the more distant it is, and
in more distant regions we can observe only a few of the brightest
galaxies.

Clusters of galaxies lie at the intersections
of filaments. Several authors have unambiguously detected 
filaments as inter-cluster structures. 
Using weak lensing, \cite{Gray} found a filament connecting
Abell clusters A901 and A902, while \citet{Dietrich}, with a similar
technique, found a filament between A222 and A223. 
In a recent paper  \citet{Pimbrecent} studied
the frequency and the distribution of filaments in 
the 2dF galaxy redshift 
survey. They were motivated by the analysis of the 
filamentary patterns of the $\Lambda$-CDM simulations reported
by \citet{Colberg}. Other observational efforts
have concluded with the detection of inter-cluster filaments 
at various redshifts \citep{Pimbblet, Ebeling}.

Although the filaments are prominent, there is no good method to
describe such a filamentary structure. The usual second moment
methods in real space or in the Fourier space (the two-point
correlation function and power spectra) do not describe well
filamentary structures. The method that has been used most is the
minimal spanning tree (MST, see a review in \citealt{martsaar02}). The
first application of the MST formalism to describe the filamentary
networks of galaxy maps was that of \citet{barrow85}; many later
studies have used it.

The minimal spanning tree is unique for a given point set, which is
good, and it connects all the points, which is not good. When the
number of galaxies is large, the MST is rather fuzzy, and it
describes mainly the local nearest-neighbour distribution (we shall
show an example of a minimal spanning tree in Sect. 5 and in 
Fig.~\ref{fig:mst}). The
filamentary network seen by eye combines both local and large-scale
features of the point distribution. Moreover, it is well known that
in the Sloan Digital Sky Survey (SDSS) and in the Two-Degree Field
Galaxy Redshift Survey (2dFGRS), there are many missing
galaxies that have not been targeted by the surveys because of a variety
of selection effects \citep{Pimb01,Cross04}. For incomplete samples
the MST is not a good choice for analysing filaments.

Thus, a better notion would be
that of the skeleton, proposed recently to describe continuous
density fields \citep{novikov03}. The skeleton is formed by lines
parallel to the gradient of the field, which connect the saddle
points to local maxima of the field. Calculating the skeleton,
however, involves smoothing the point distribution, which will
introduce an extra parameter, therefore this method is not well
suited for point distributions.

We propose to use an automated method to trace filaments for
realizations of point processes that has been shown to work well
for the detection of road networks in remote sensing 
\citep{LacoDescZeru02,StoiDescLiesZeru02,StoiDescZeru04}. This method
is based on the Candy model, a marked point process, where segments
serve as marks. As this is the first time such a method is used for
the galaxy distribution, we describe it in detail below. We also test it
on 2-D simulated galaxy maps, justifying our model choice. The
task differs considerably from road network detection, as the noise
is larger, and we have no continuous roads, but sparsely populated
ridges instead.

The present approach allows us to find the length distribution for
the filaments; we give examples of this distribution for different
data samples. In this paper, we choose the Candy process parameters
by trial and error following a reversible jump process. As the
method is automated, it can also be used to estimate those
parameters by using maximum likelihood methods; these will serve
then as new statistics for filament networks.

\section{Marked point processes}
Let $(K,\mathcal{B},\nu)$ be a measure space, where $K$ is a
compact subset of $\R^{2}$ of strictly positive Lebesgue measure
$0< \nu(K) < \infty$ and $\mathcal{B}$  the associated Borel
$\sigma-$algebra of subsets of $K$. For $n \in \N$ let $K_{n}$ be
the set of all unordered configurations
$\kk=\{k_1,k_2,\ldots,k_n\}$ that consist of $n$ not necessarily
distinct points $k_i \in K$. Let us consider the configuration
space $\Omega = \cup _{n=0}^{\infty}K_n$ equipped with the
$\sigma-$algebra $\mathcal{F}$ generated by the mappings
$\{k_1,k_2,\ldots,k_n\}\rightarrow \sum_{i=1}^{n}\1\{k_i \in B\}$
counting the number of points in Borel sets $B \in \mathcal{B}$. A
point process on $K$ is a measurable map from a probability space
into $(\Omega,\mathcal{F})$. For introductory material on
point processes we refer the reader to the textbooks by 
\citet{Lies00} and \citet{reiss}. 

The reference measure is given by the unit rate Poisson process
that distributes the points in $K$ according to a Poisson process
with intensity $\nu$.

Different characteristics or marks may be attached to the points.
Under these circumstances, we consider a point process on $K \times
M$ as the random sequence $\xx=\{(k_1,m_1),\ldots,$
$(k_n,m_n)\}$ where
$n \in \N_{0}$, $k_i \in K$ and $m_i \in M$ for all $i=1,\ldots,n$.
The characteristics space $M$ is equipped with its corresponding
Borel $\sigma-$algebra and the probability measure $\nu_M$. A marked
point process $X$ with locations in $K$ and marks in $M$ is a point
process on $K \times M$ such that the distribution of locations only
is a point process on $K$.

In this case, the reference measure is the unit rate Poisson process
on $K \times M$, with the locations distributed according to a
Poisson process with intensity $\nu$ and i.i.d\footnote{i.i.d. -- independent
and identically distributed} marks according to
$\nu_{M}$. When the marks represent parameters of an object, such a
process is sometimes called an object point process.

The reference measure exhibits no interaction between points or
objects. Indeed, we can construct a much more complicated marked
point process by specifying a probability density with respect to
the reference measure:
\begin{equation}
p(\xx)=\alpha \exp[-U(\xx)], \label{gibbs_distribution}
\end{equation}

\noindent with $\alpha$ the normalizing constant and $U(\xx)$ the
interaction energy of the system. The energy function is written
as the sum
\begin{equation}
U(\xx)=\sum_{j=1}^{q}\sum_{\{x_{i1},\ldots,x_{ij}\}\subseteq
\xx}\omega^{(j)}(x_{i1},\ldots,x_{ij})
\end{equation}

\noindent where $\xx$ is a realization of the marked point process,
$\omega^{(j)} : (K \times M)^{j} \rightarrow \R$ for
$j=1,\ldots,q$ are the interaction potentials. The marked point
processes with a probability density of the form given by
Eq.~(\ref{gibbs_distribution}) are known in physics under the name of
Gibbs point processes. If there exists a positive real $C
> 0$ such that $U(\xx)-U(\xx \cup \{(k,m)\}) \leq \log C $ for all
$(k,m) \in K \times M$ the process is said to be locally stable.

This relation implies the Ruelle stability condition
\citep{Ruel69}, which ensures the integrability of a given
probability density function. Furthermore, local stability is
essential in establishing convergence proofs for the Monte Carlo
dynamics simulating such a model \citep{Geye99}.

For our problem, $\yy$, the data to be analysed, consist of
points (galaxies) spread in a finite window $K$. We want to
extract the filamentary structure of these data. It is natural to
consider the filaments $\xx$ we want to detect as a set of random
segments being the realization of a marked point process.

The probability density of such a marked point process is given by
\begin{equation}
p_{\yy}(\xx) \propto \exp[-(U_{\yy}(\xx)+U_{r}(\xx))]
\label{gibbs_data_distribution}
\end{equation}

\noindent with the terms $U_{\yy}(\xx)$ and $U_{r}(\xx)$ being the
data energy and the interaction energy, respectively. The first
term is related to the location of the filaments among the galaxies,
whereas the second is related to the geometrical properties of the
filaments, playing the role of a regularization term.

The configuration of segments composing the filamentary network is
estimated by the minimum of the total energy of the system
\begin{equation}
\hat{\xx}=\arg\min_{\xx}\{U_{\yy}(\xx)+U_{r}(\xx)\}.
\label{estimate_network}
\end{equation}

In the following we will present the two components of the energy
function. Considerations about the simulation of such models using
the Monte Carlo Markov  Chain dynamics will be given and a simulated annealing algorithm
will be presented. Finally, we will apply the model to describe
two-dimensional filamentary networks of galaxies.

\section{A probabilistic model for the filamentary structure of galaxy maps}

\subsection{The interaction energy: Candy model}
The filaments we want to extract are composed of non-overlapping
connected segments. Locally, the curvature of one filament does not
vary too much. In our data we can notice just a few short
filaments, which can be represented by isolated segments.

Under these considerations a natural choice for the interaction
energy becomes the Candy model, a marked point process simulating
random networks of segments. Here, a segment is seen as a random
object $\zeta=(k,(\theta,w,l))$ that is characterized by its center
location $k \in K$ and its geometrical parameters $(\theta,w,l) \in
[0,\pi] \times [w_{\min},w_{\max}] \times [l_{\min},l_{\max}]=M$,
representing its orientation, width and length respectively. The
Candy model exhibits three types of interactions between segments:
connectivity, alignment and rejection.

Historically speaking, the model was introduced for the first time
as a prior distribution for thin network extraction in remotely
sensed images \citep{Stoi01,StoiDescLiesZeru02,StoiDescZeru04}.
Properties of the model such as local stability and Markovianity,
convergence proofs of an adapted Metropolis-Hastings dynamics for
simulating the model as well as parameter estimation were further
investigated in \citet{LiesStoi03}. Different versions of the model
were analysed and compared for the special case of road network
detection \citep{LacoDescZeru02}.

A segment has a connection region formed by the union of the two
circles centered at its extremities and of radius $r_c$. Two
segments $\eta=(k_\eta,(\theta_{\eta},w_{\eta},l_{\eta}))$ and
$\zeta=(k_{\zeta},(\theta_{\zeta},w_{\zeta},l_{\zeta}))$ are
connected $\eta \sim_{c} \zeta$ if only one extremity of a segment
is in the connection region of the other segment and if
$\|\theta_{\eta}-\theta_{\zeta}\| \leq \tau$. With respect to this
definition, a segment is doubly connected if both of its
extremities are connected, singly connected if only one of its
extremities is connected and free if none of its extremities is
connected. The Candy model favours doubly connected segments
whereas free and singly connected segments are penalized.

\begin{figure}
\centering
\resizebox{0.45\textwidth}{!}{\includegraphics*{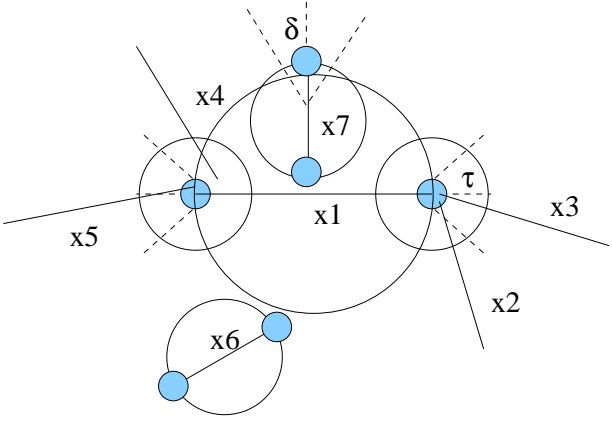}}
\caption{Connection and alignment interaction between segments.
The circle around a segment represents the rejection region, 
whereas the circles around its extremities indicate the connection 
region. The values $\tau$ and $\delta$ are the curvature values 
allowing segments to align and to cross, respectively. 
\label{fig:cntsegments}}
\end{figure}

In Fig.~\ref{fig:cntsegments} we show an example of a configuration
of segments. The free segments are $x_2,x_4,x_6$ and $x_7$, this
because the segment $x_2$ does not fullfil the orientation
requirements for the connection and the others do not respect the
connection condition. The segments $x_3$
 and $x_5$ are singly connected, whereas the segment $x_1$ is
 doubly connected.

Similarly, the attraction region of a segment $\eta$ is the union of
both circles centered at each extremity with a radius
$r_{o}=l_{\eta}/4$. Two segments $\eta$ and $\zeta$ exhibit
alignment interaction $\eta \sim_{o} \zeta$ if
$d(k_{\eta},k_{\zeta})
> \frac{1}{2}\max\{l_{\eta},l_{\zeta}\}$, if only one extremity of
a segment is in the attraction region of the other segment, and if
$\min\{\|\theta_{\eta}-\theta_{\zeta}\|,\pi-\|\theta_{\eta}-\theta_{\zeta}\|\}
> \tau$, with $\tau$ a threshold value. The Candy model penalizes the
segments having alignment interaction.

In the configuration shown in Fig.~\ref{fig:cntsegments} $x_3
\not\sim_{o} x_1$ and $x_5 \not\sim_{o} x_1$, while the segments
$x_2 \sim_{o} x_1$ and $x_4 \sim_{o} x_1$ because these pairs of
segments exhibit high differences between their orientations.

Connectivity is a stronger interaction than alignment. Still, as
we look for the filaments fitting the data in a random way, this
last interaction gives us the possibility not to eliminate from
the current configuration the segments with low data energy, which
are almost connected.

Every segment $\eta$ is provided with a rejection region given by
a circle centered in $k_{\eta}$ and of a radius $r_r=l_{\eta}/2$.
Two segments $\eta$ and $\zeta$ exhibit rejection interaction if
$d(k_{\eta},k_{\zeta})<\frac{l_{\eta}+l_{\zeta}}{2}$ and if
$\|\|\theta_{\eta}-\theta_{\zeta}\|-\pi/2\|>\delta$, where
$\delta$ is a threshold value. The Candy model forbids
configurations containing rejecting (overlapping) segments.

If $d(k_{\eta},k_{\zeta}) \leq
\frac{1}{2}\max\{l_{\eta},l_{\zeta}\}$ and if
$\|\|\theta_{\eta}-\theta_{\zeta}\|-\pi/2\| \leq \delta$, then the
segments may cross or form a ``T" junction. The configurations with
crossing segments $\eta \sim_{x} \zeta$ are forbidden by the Candy
model, whereas the ``T" junctions are allowed.

Clearly, in Fig.~\ref{fig:cntsegments} the segments $x_1$ and $x_6$
do not reject each other since they are far enough apart, while the
segments $x_1$ and $x_7$ do not cross, forming a "T" junction.

For any configuration of segments $\xx=\{x_1,\ldots,x_n\}$ with
$i=1,\ldots,n$, we are able now to write for the probability
density of the Candy model
\begin{equation}
\begin{array}{ccl}
p_{r}(\xx) & \propto & \left \{ \prod_{i=1}^{n(\xx)} \exp \left[
\frac{l_i-l_{\max}}{l_{\max}} +
\frac{w_i-w_{\max}}{w_{\max}}\right]  \right \} \times\\
& &
\gamma_{d}^{n_{d}(\xx)}\gamma_{f}^{n_{f}(\xx)}\gamma_{s}^{n_{s}(\xx)}\gamma_{o}^{n_{o}(\xx)}
\times\\
& &
 \prod_{i<j}\1\{x_i \not\sim_r x_j\}\1\{x_i \not\sim_{x} x_j\}
\end{array}
\label{candy_density}
\end{equation}

\noindent where $\gamma_d,\gamma_f,\gamma_s >0$ and $\gamma_o \in
(0,1)$ are the model parameters, $n_d(\xx),n_f(\xx),n_s(\xx)$ are
the numbers of doubly, free and singly connected segments, and
$n_o(\xx)$ is the number of pairs of segments which are not well
aligned. In order to avoid too many small segments in the
configuration, the model favours segments covering a large area.
Clearly the interaction energy is obtained taking $U_{r}(\xx)=-\log
p_{r}(\xx)$.

With respect to the classical definition of the Candy model in
\citet{LiesStoi03}, the model described by Eq.~(\ref{candy_density})
contains differences in the definition of interactions between
segments. We kept the same name for our model, as we believe that
the modifications required to apply it to cosmological data do not
change the basic premises of the classical Candy model. Concerning
connectivity, the present modifications were introduced in order to
eliminate some ``undesired" configurations, such as a segment being
connected with itself or a segment being connected at one extremity
with both extremities of another segment. Furthermore, the new
modifications allow us to build more appropriate tailored moves for
the Metropolis-Hastings dynamics simulating the model. The rejection
region was extended, as the filaments we observe may be rather wide,
hence we want to avoid overlapping of segments when the data are good
enough. This is also the reason why the width penalty was
introduced. Nevertheless, it is easy to prove that under these
modifications, together with the one concerning the 
crossing interaction the Candy
model is still locally stable and (Ripley-Kelly) Markovian
\citep{RiplKell77}.

\subsection{The data energy}
\begin{figure}
\centering
\resizebox{.4\textwidth}{!}{\includegraphics*{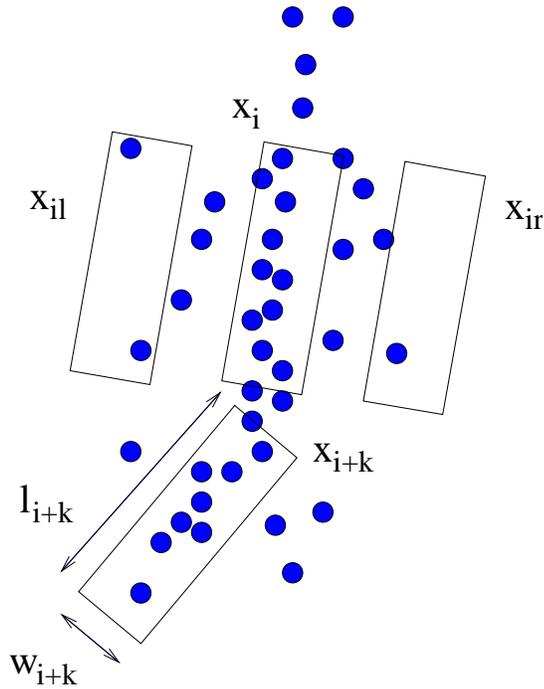}}
\caption{Locating segments in a pattern of points.
\label{fig:datasegments}}
\end{figure}

The data energy checks whether a segment belongs to the network or
not \citep{Stoi01,StoiDescLiesZeru02,StoiDescZeru04}. A segment $x$
is considered a part of the filament network, if its geometrical
shape $\tilde{x}$ covers as many galaxies as possible. Still, we
want to avoid the case where segments are found in a cloud of points
rather than in a filament. Clouds
of points can also be considered as inter-cluster filaments, but
we want to favour the selection of the shoelace-like filaments,
which are the more common morphological type \citep{Colberg,Pimbblet},
although other shapes (ribbons and sheets) can still be detected
by the Candy model.  
To do this, we consider the shadow
segments $x_{r}$ and and $x_{l}$ -- the segments situated to the
right and to the left of the segment $x$, as in
Fig.~\ref{fig:datasegments}. The above-mentioned case is avoided if
the number of galaxies covered  by $\tilde{x_{r}}$ and
$\tilde{x_{l}}$ is small. Therefore, let us define the quantity
$v_{\yy}(x)$ given by
\begin{equation}
v_{\yy}(x)=2n(\yy \cap \tilde{x})-n(\yy \cap \tilde{x_{r}})-n(\yy
\cap \tilde{x_{l}}),
\end{equation}

\noindent where $n(\yy \cap \tilde{x})$ is the number of galaxies
covered by the geometrical shape of the segment $x$. Now, if the
following three conditions: $v_{\yy}(x) \geq 3$, $n(\yy \cap
\tilde{x}) > n(\yy \cap \tilde{x_{r}})$, and $n(\yy \cap \tilde{x})
> n(\yy \cap \tilde{x_{l}})$  are simultaneously fulfilled, the
data energy contribution of a segment is
$V_{\yy}(\{x\})=-v_{\yy}(x)$. If one of the three conditions
is not fulfilled then $V_{\yy}(\{x\})=V_{\max}$, with $V_{\max} >
0$ a positive fixed value.

The total data energy is defined as the sum of the data energy
contributions of every segment in the configuration
\begin{equation}
U_{\yy}(\xx)=\sum_{x \in \xx} V_{\yy}(\{x\}) \label{data_energy}
\end{equation}

\subsection{Simulation dynamics and optimization}
The equations (\ref{candy_density}) and (\ref{data_energy})
provide us with all the ingredients needed to construct the Gibbs
point process  given by Eq.~(\ref{gibbs_data_distribution}). The
estimate of the network (Eq.~\ref{estimate_network}) is obtained by
means of a simulated annealing algorithm.

This algorithm iteratively samples the law
$[p_{\yy}(\xx)]^{\frac{1}{T}}$ while slowly decreasing the
temperature $T$. At high positive temperature values the 
state space is explored. When the temperature goes down to zero, $T
\rightarrow 0$, the configurations of minimal energy are reached.
A polynomial decreasing scheme $T_{k+1}=cT_{k}$ with $c \in
[0.99,1.00]$ may be used for cooling.

For sampling from a probability density of a point process several
Monte Carlo methods are available, such as the spatial
birth-and-death process, the Metropolis-Hastings and reversible
jumps dynamics, or the much more recent exact simulation
techniques such as coupling from the past or clan of ancestors
\citep{GeyeMoll94,Geye99,Gree95,KendMoll00,Lies00,LiesStoi03a,Pres77}.
The Candy point process exhibits rather complicated interactions,
hence the use of the spatial birth-and-death process or the cited
exact simulation techniques are useful in practice only for a
limited range of the model parameters. Therefore, for our present
model we opted for a sampling algorithm based on the
Metropolis-Hastings dynamics. Details concerning the
implementation of samplers for the Candy model based on
Metropolis-Hastings or reversible jumps processes can be found in
\citet{LiesStoi03,Stoi01,StoiDescLiesZeru02,StoiDescZeru04}.

\section{Data}

\begin{figure}
\centering
\resizebox{0.45\textwidth}{!}{\includegraphics*{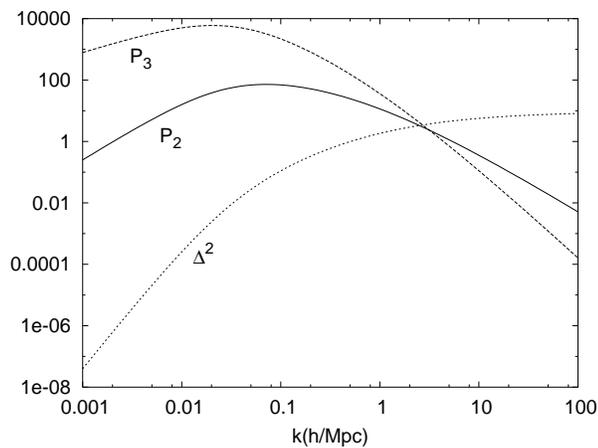}}
\caption{The spectral density used for the 2-D simulation ($P_2$),
the corresponding spectral density for the 3-D case ($P_3$), and
the spectral energy per unit logarithmic wavenumber interval
$\Delta^2$, versus the wavenumber $k$. \label{fig:power}}
\end{figure}

\begin{figure}
\centering
\resizebox{!}{0.32\textwidth}{\includegraphics*{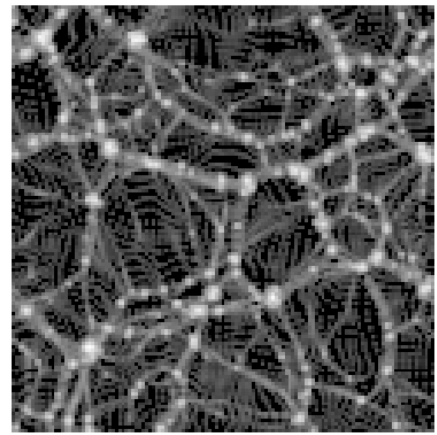}}\\
\caption{Dark matter density (logarithmic scale) for the data set B.
\label{fig:densB}}
\end{figure}

\begin{figure*}
\centering
\resizebox{!}{.32\textwidth}{\includegraphics*{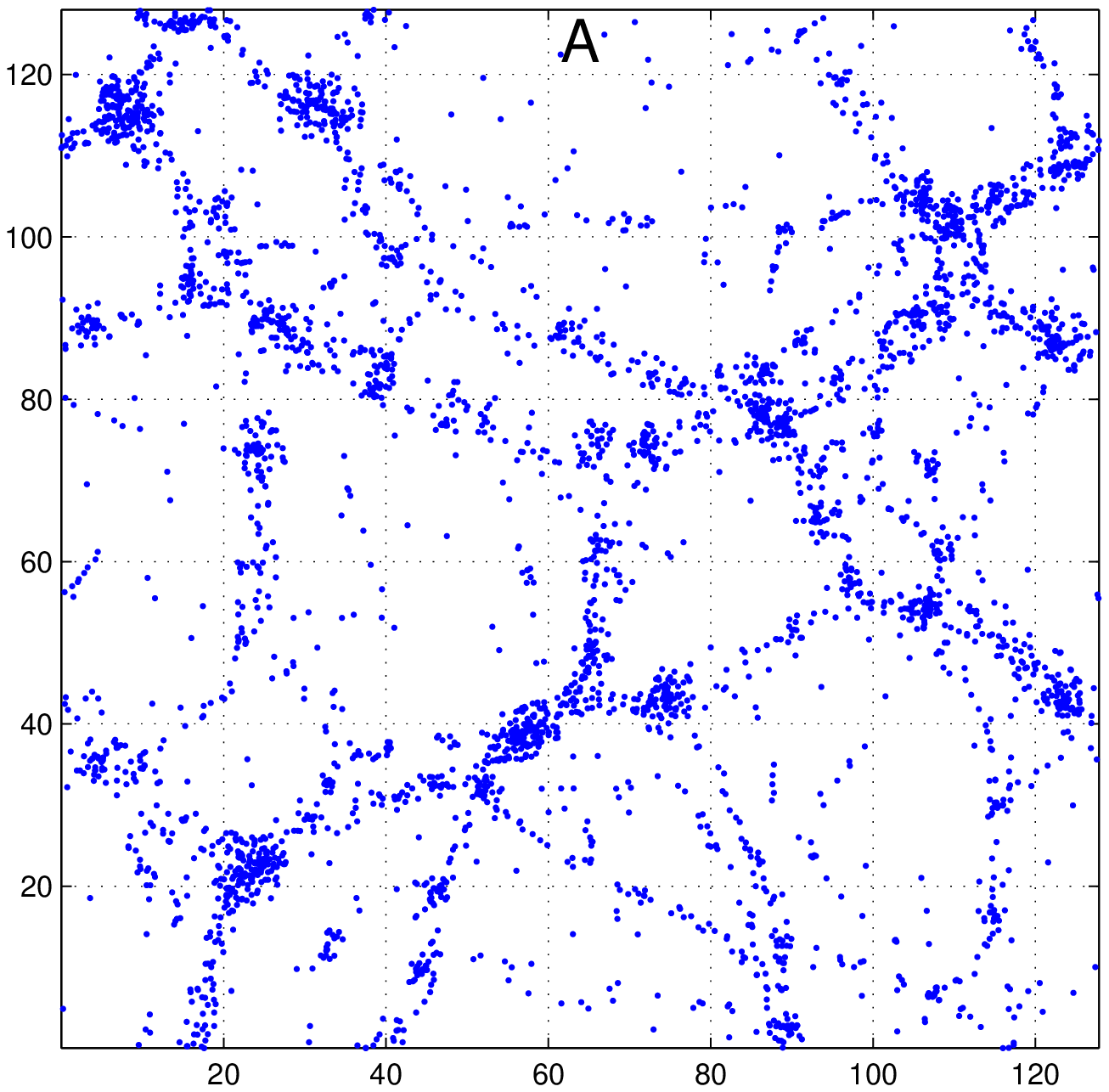}}
\resizebox{!}{.32\textwidth}{\includegraphics*{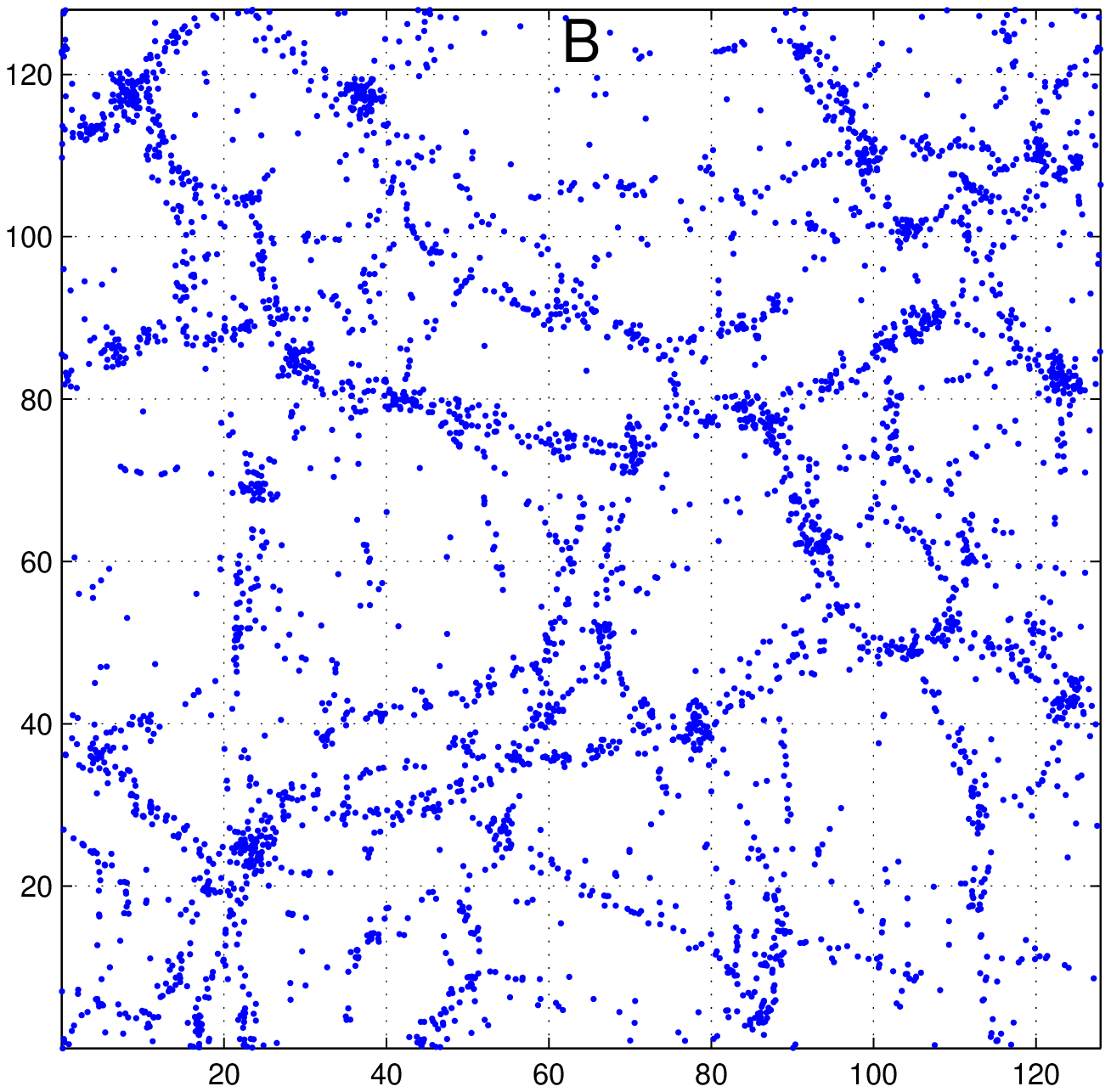}}
\resizebox{!}{.32\textwidth}{\includegraphics*{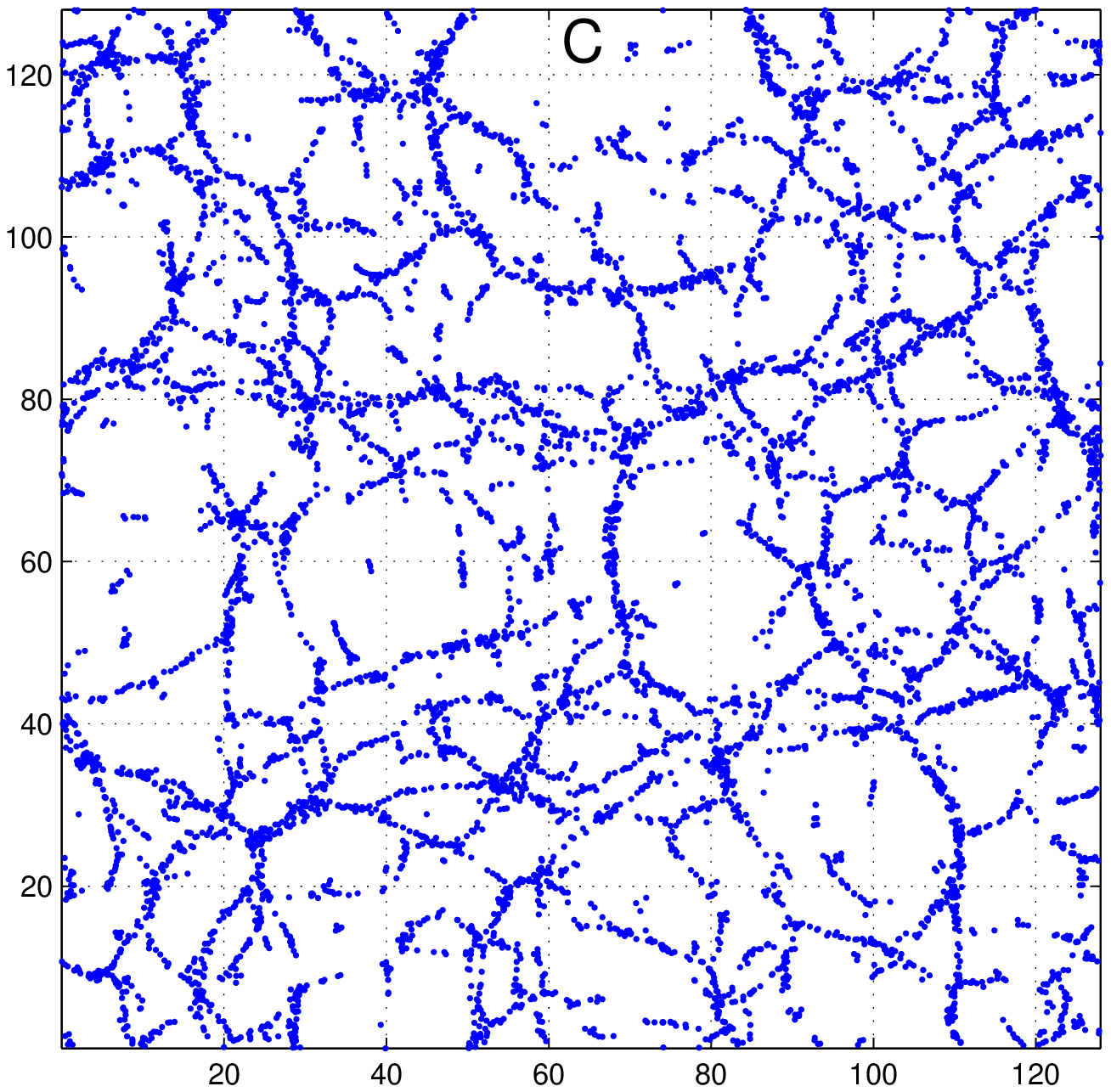}}\\
\caption{The simulated galaxy distribution for the three
sets of data. Panel B corresponds to the dark matter distribution
shown in Fig.~\ref{fig:densB}. In this and the following figures
the units of distance are $h^{-1}$Mpc, though 
in a 2-D world. 
\label{fig:alldata}}
\end{figure*}

\begin{table}
\centering \caption{Parameters of the data sets: $a$ is the
cosmological expansion factor, $n$ is the number of galaxies,
$\alpha$ is the void density threshold and $\beta$ is the biasing
amplitude. \label{tab:data}} \vspace*{2pt}
\begin{tabular}{c|cccc}\hline
Case&$a$&$n$&$\alpha$&$\beta$\\
\hline
A&1.0&4127&0.5&0.20\\
B&0.6&4249&0.5&0.18\\
C&0.2&8879&1.0&0.49\\
\hline
\end{tabular}
\end{table}

The Candy process and its applications have been developed for 2-D
maps. So the natural way to introduce them in cosmology is to
consider 2-D cases, also. This will allow us to compare the results,
and will make it easier to understand the problems arising. Our
final goal is to apply the Candy process to describe 3-D networks
of filaments, as the large-scale structure maps fill the space.
The 3-D network consists of complete filaments, as do the 2-D
geographical road maps, so the filaments in the test data should
also be complete.

The observational galaxy maps showing filaments 
mainly have the geometry of a thin slice. 
Although such data have been used to
study the large-scale filamentary structure, the slices do not
provide proper data for that. The thickness of these slices is much
smaller than the typical size of a filament, and although the maps
give a visual impression of filaments, the filaments we see are
pieces of real filaments, obtained by planar cuts through the real
3-D structure.

    Another possibility is to use thicker slices,
which can be selected, e.g., from the only large-scale contiguous
data set for the moment, the 2dF survey \citep{colless03}. But this
choice carries its own difficulties -- thick slices give us the 2-D
projection of the 3-D network, smearing essential details.

Simulations of the formation and evolution of large-scale structure
can also provide us with galaxy maps. As a demonstration that we
understand the basic features of the process, these maps show
filamentary structure. How and why an initial Gaussian random
density field develops filaments under self-gravitation is 
well explained by \citet{bond96}.

The usual simulations give us 3-D universes, but it is easy to also
simulate the evolution of structure in a 2-D universe. This has been
done before, to obtain better numerical resolution (see, e.g.,
\citealt{beacom91}); we used 2-D simulations to get complete
cosmological networks of model galaxies.

The present-day large-scale structure is determined, first, by the
expansion history of the cosmological model, and, second, by the
initial density and velocity fields at the start of the
simulation. We chose the standard 'concordance' cosmological model
\citep{tegmark01} to describe the expansion. As the initial fields
are assumed to be Gaussian random fields, they are described by
their power spectra (the spectral density of the density
perturbations $P(k)$, where $k$ is the module of the wave-vector;
see, e.g., \citealt{martsaar02}). We chose a simple expression for the
spectral density that describes reasonably well the Cold Dark
Matter (CDM) model \citep{jenkins98} and modified it to get the same
spectral energy contribution to the variance  per unit logarithmic
wavenumber interval, $\Delta^2(k)$, in our 2-D universe as in the
real 3-D universe. In a 3-D universe this quantity is defined as
\[
\Delta^2_3(k)=\frac{1}{2\pi^2}P_3(k)k^3,
\]
and in a 2-D universe as
\[
\Delta^2_2(k)=\frac{1}{2\pi}P_2(k)k^2;
\]
the equality of the above quantities gives
\[
P_2(k)=\frac{k}{\pi}P_3(k)
\]
(the lower indices show the dimensionality of the space). This is
the spectral density we used, with $P_3(k)$ taken from
\citet{jenkins98}. Both the spectral densities and the spectral
energy used are shown in Fig.~\ref{fig:power}. As usual, the
wavenumber is given in units of $h/\mbox{Mpc}$
where $h$ is the
dimensionless Hubble parameter, the spectral
densities are in units of Mpc$^3/h^3$ ($P_3(k)$), Mpc$^2/h^2$
($P_2(k)$), and $\Delta^2(k)$ is dimensionless.

We selected the scales and spectrum amplitudes to get a picture
similar to that we see in 3-D models (the size of the patch we
modelled was $128h^{-1}$Mpc, and we used a $256^2$ grid with the
same number of cold dark matter particles). These numbers are not
really important, as this is a mock model, anyway. Then we ran a 2-D
dynamical $N$-body simulation, developing the initial perturbations
into large-scale structures -- the present-day density and velocity
fields.

These density fields describe the dark matter content of the
universe. Populating model universes with galaxies is a complex
problem, but for our purposes simple recipes are sufficient. We
used two well-known properties of the large-scale galaxy
distribution. First, galaxies avoid large low-density regions,
known as voids; we modeled this by selecting a density threshold
$\alpha$ (all our densities are given in the units of the mean
density). In regions with density lower than this threshold no
galaxies were placed. Secondly, galaxy density is biased in
respect to the dark matter density. We found that the model galaxy
distribution best resembled the observational maps for a nonlinear
biasing law:
\begin{equation}
\label{bias} \rho_{\mbox{\scriptsize
gal}}=\beta\sqrt{\rho_{\mbox{\tiny CDM}}},\qquad
    \rho_{\mbox{\tiny CDM}}\ge\alpha.
\end{equation}
We chose the amplitudes $\alpha$ and $\beta$ to produce
approximately the same number of galaxies as observed in
cosmological slices of similar size.

Finally, we generated a realization of a Cox point process, using
the galaxy density given by Eq.~(\ref{bias}) as the driving probability.
In order to see how well the Candy model works in different
situations, we chose three different time moments from the
simulation, with a different filamentary structure.  As the earliest
of them (our case C) has a very rich set of filaments, we generated
about twice as many galaxies for that data set as for other sets. As
usual in cosmology, we characterize the time moments by the value of
the expansion factor $a$. This factor equals unity at the present
epoch, and the earlier the epoch, the smaller is the expansion factor
(our universe expands). The parameters for our three data sets are
given in Table~\ref{tab:data}, and the dark matter density and
galaxy distributions are compared for the set B in
Fig.~\ref{fig:densB}.

All the data sets are shown in Fig.~\ref{fig:alldata}.

\subsection{Experimental results}

\begin{figure*}
\centering
\resizebox{0.32\textwidth}{!}{\includegraphics*{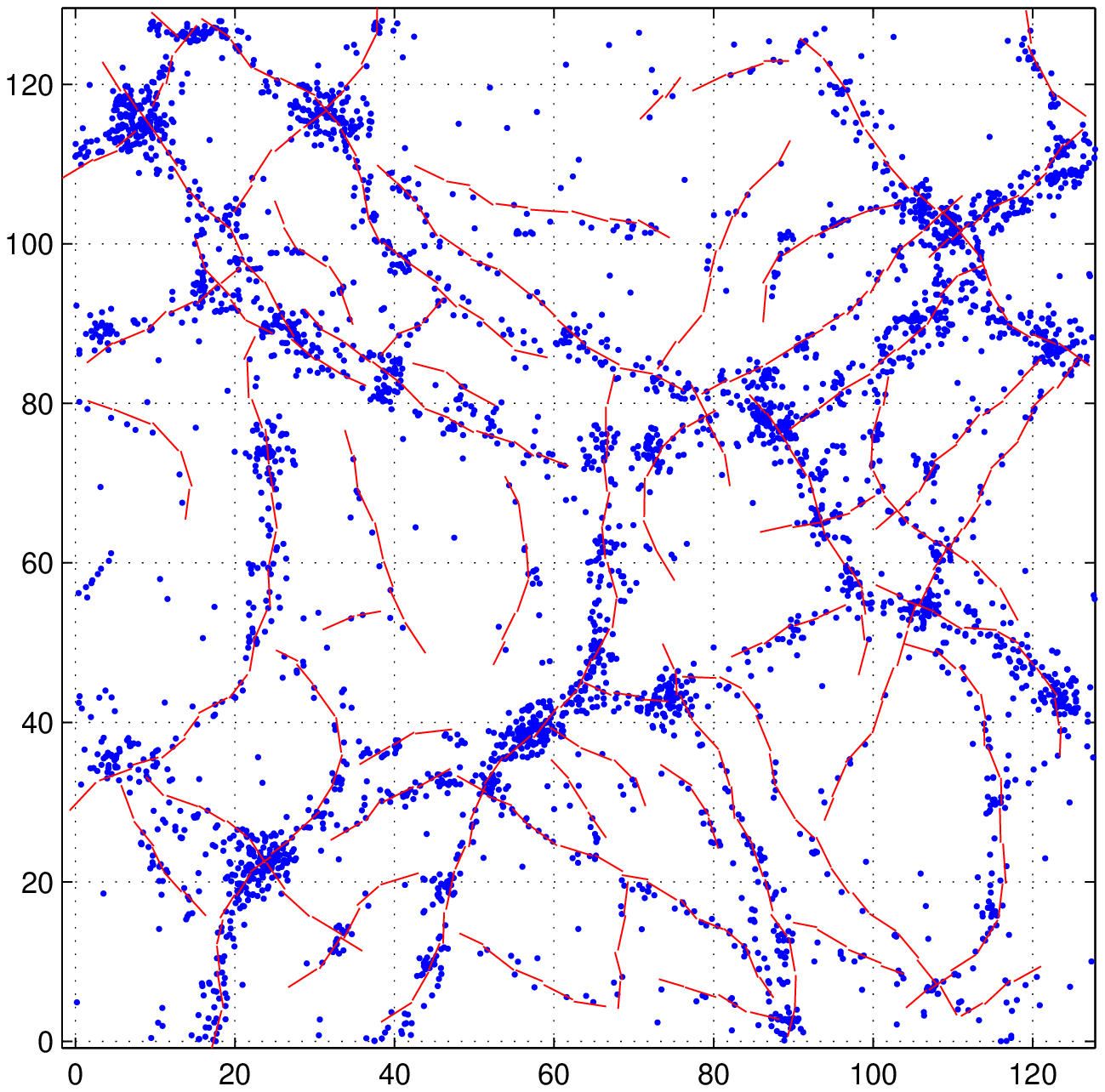}}
\resizebox{0.32\textwidth}{!}{\includegraphics*{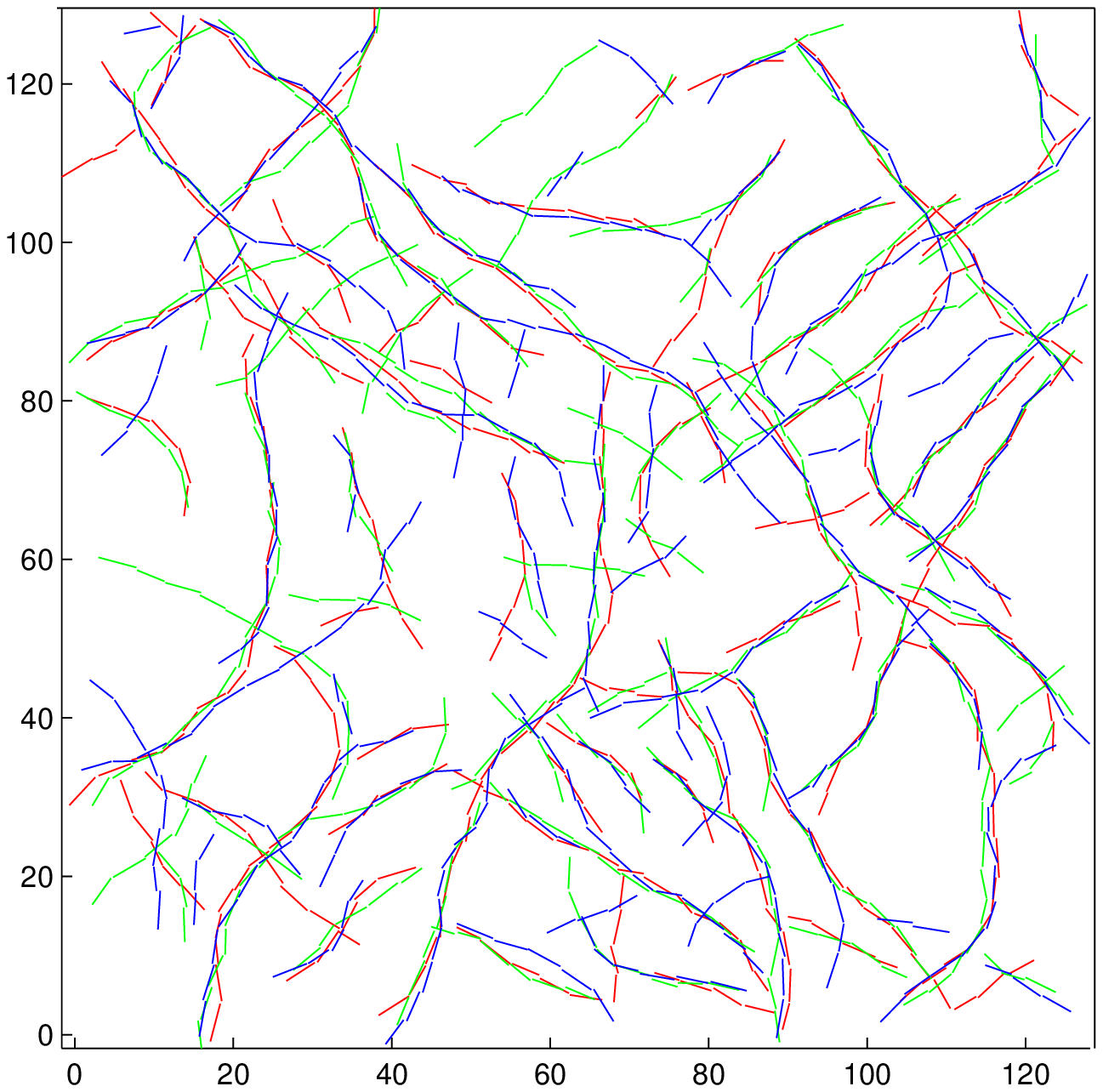}}\\
\resizebox{0.32\textwidth}{!}{\includegraphics*{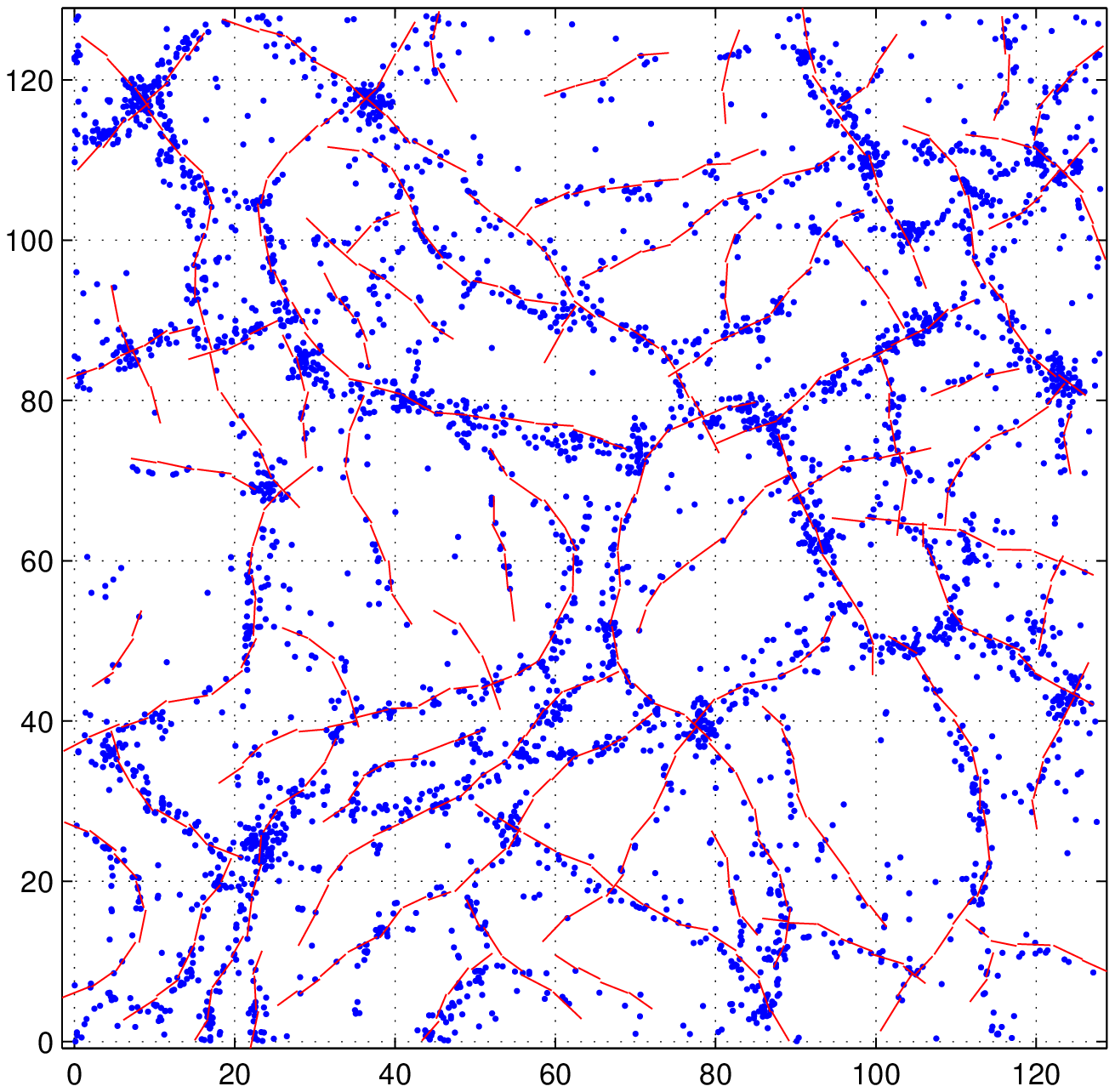}}
\resizebox{0.32\textwidth}{!}{\includegraphics*{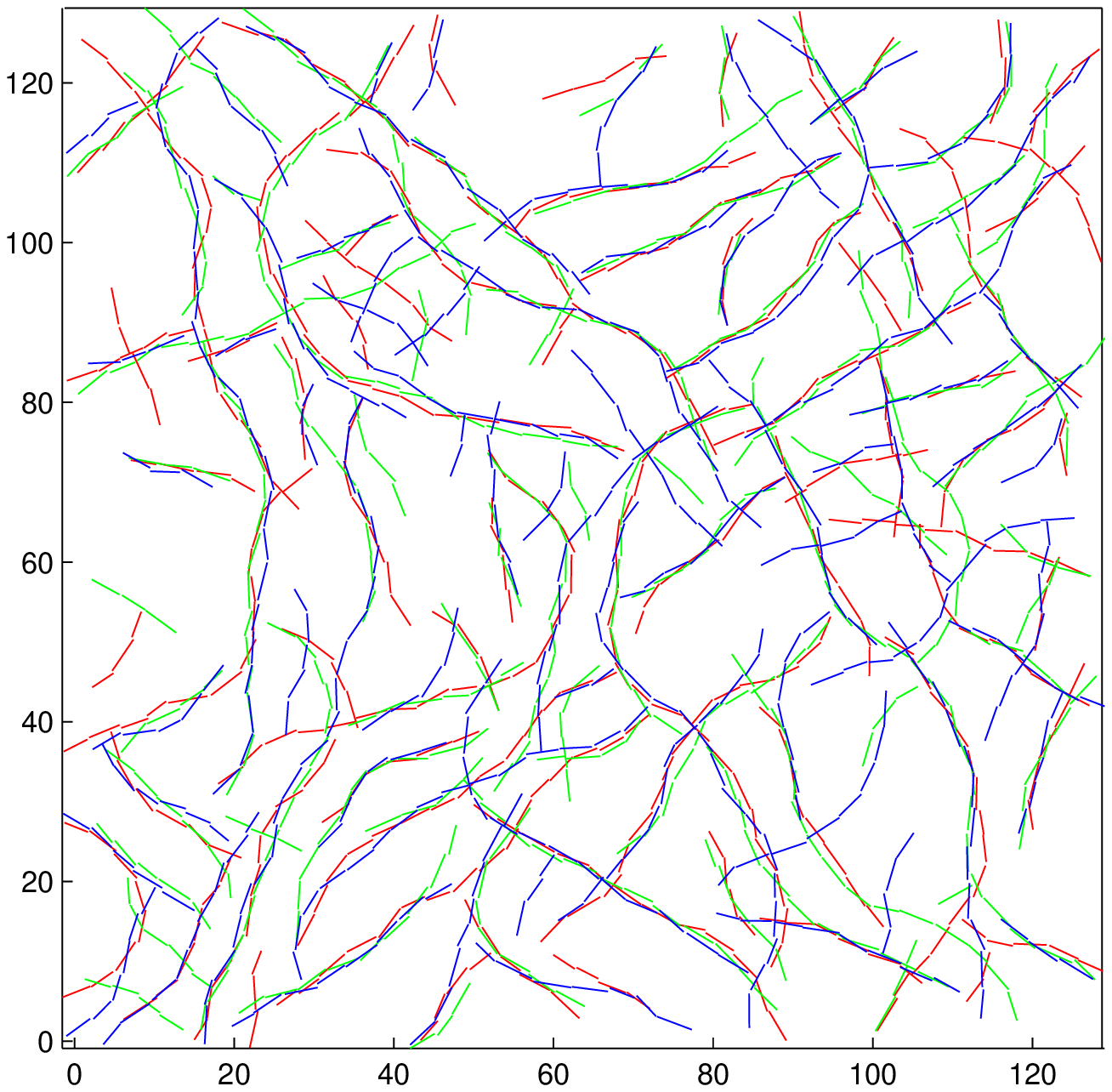}}\\
\resizebox{0.32\textwidth}{!}{\includegraphics*{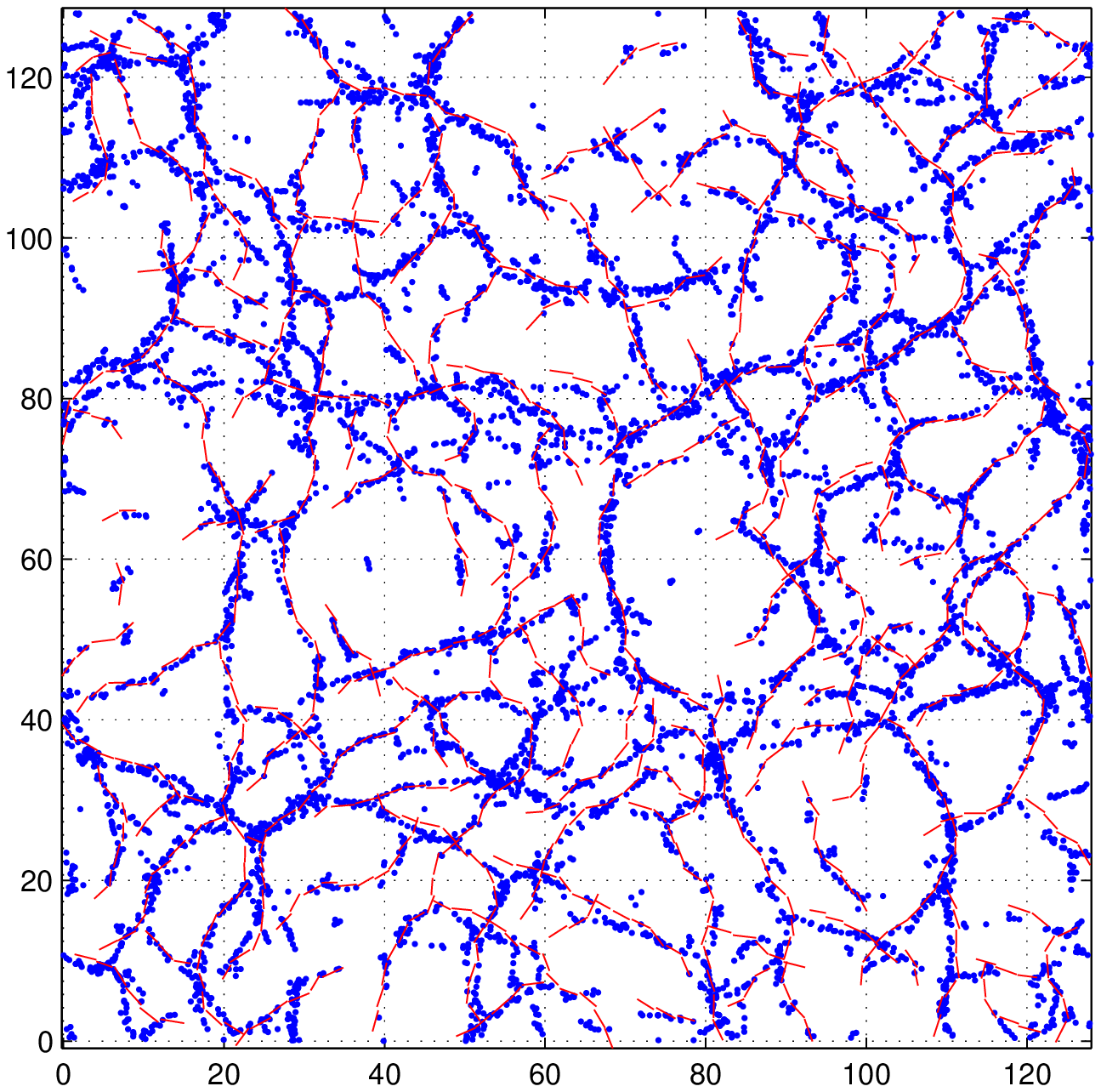}}
\resizebox{0.32\textwidth}{!}{\includegraphics*{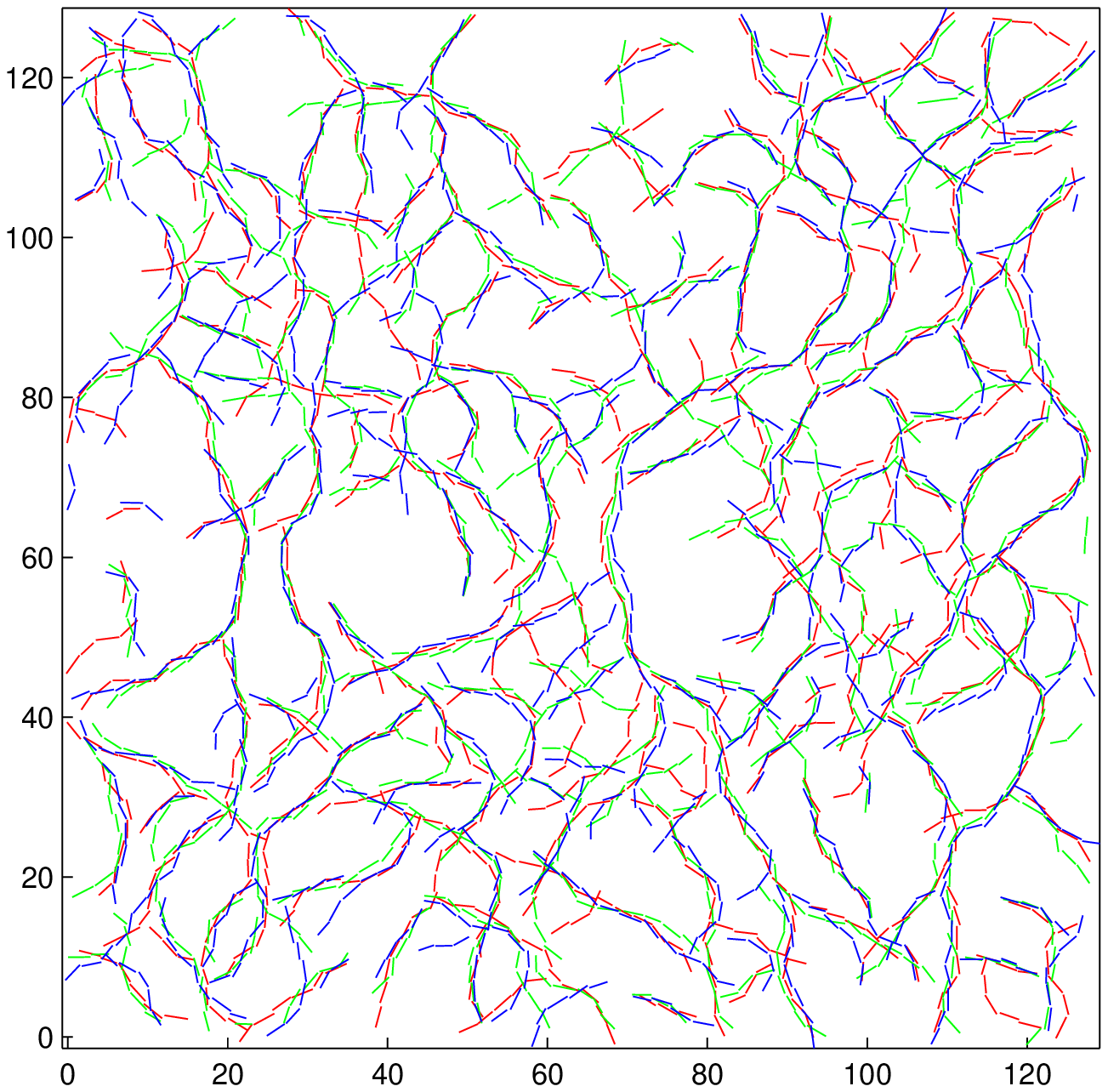}}\\
\caption{Results obtained for the three data sets: A (top row),
B (middle row), and C (bottom row). The left panels show the
``best network'' extraction superposed on the data, the right panels 
show the three networks superposed. The networks for the first
parameter set are shown by red curves, for set 2 by blue
curves, and for set 3 by green curves. \label{fig:ds1}}
\end{figure*}
\begin{figure*}
\centering
\resizebox{0.45\textwidth}{!}{\includegraphics*{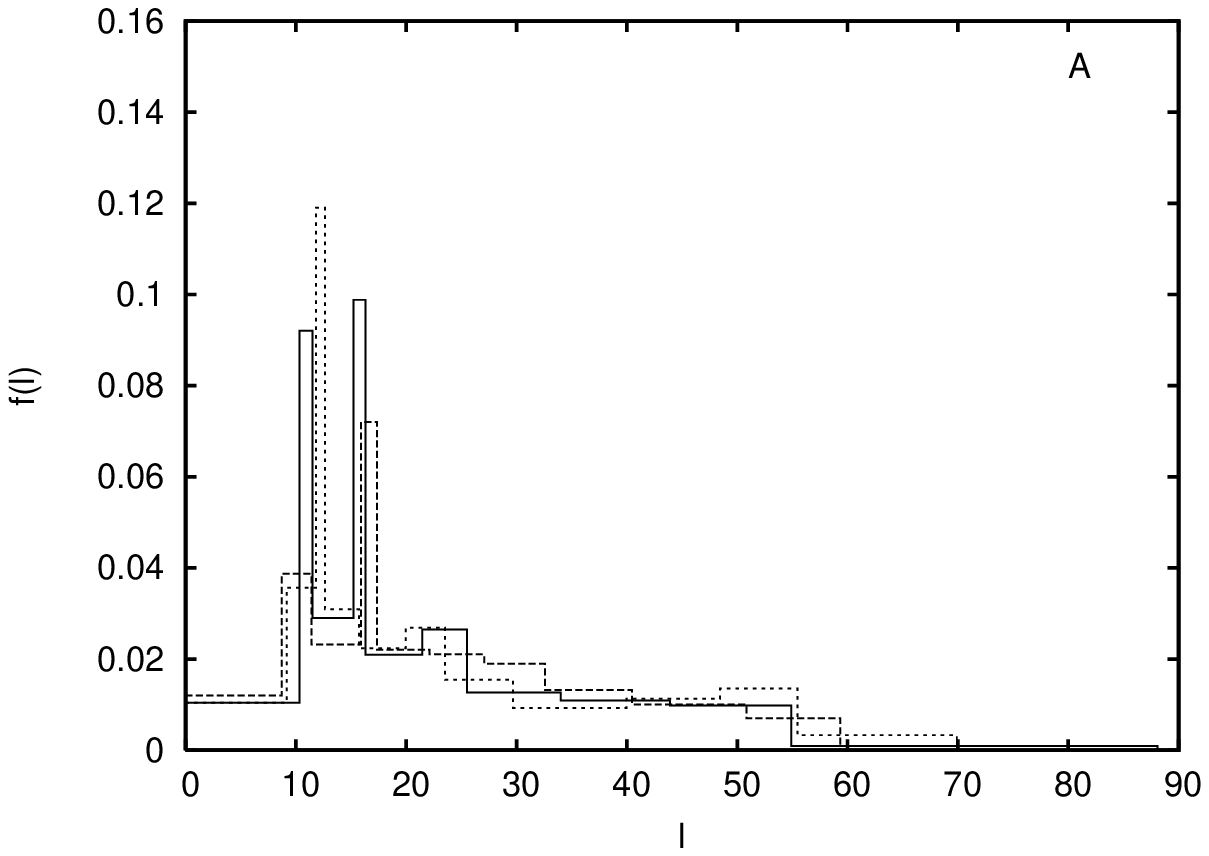}}
\resizebox{0.45\textwidth}{!}{\includegraphics*{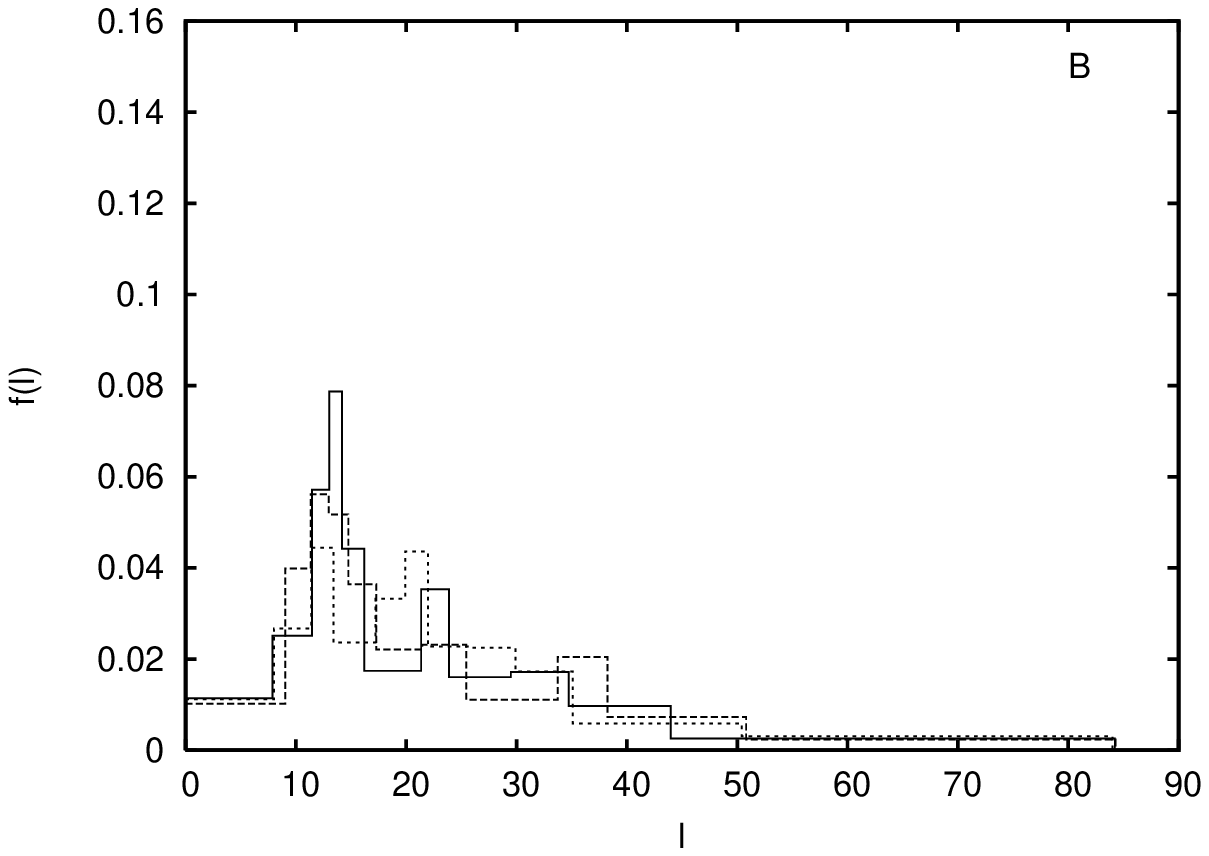}}
\resizebox{0.45\textwidth}{!}{\includegraphics*{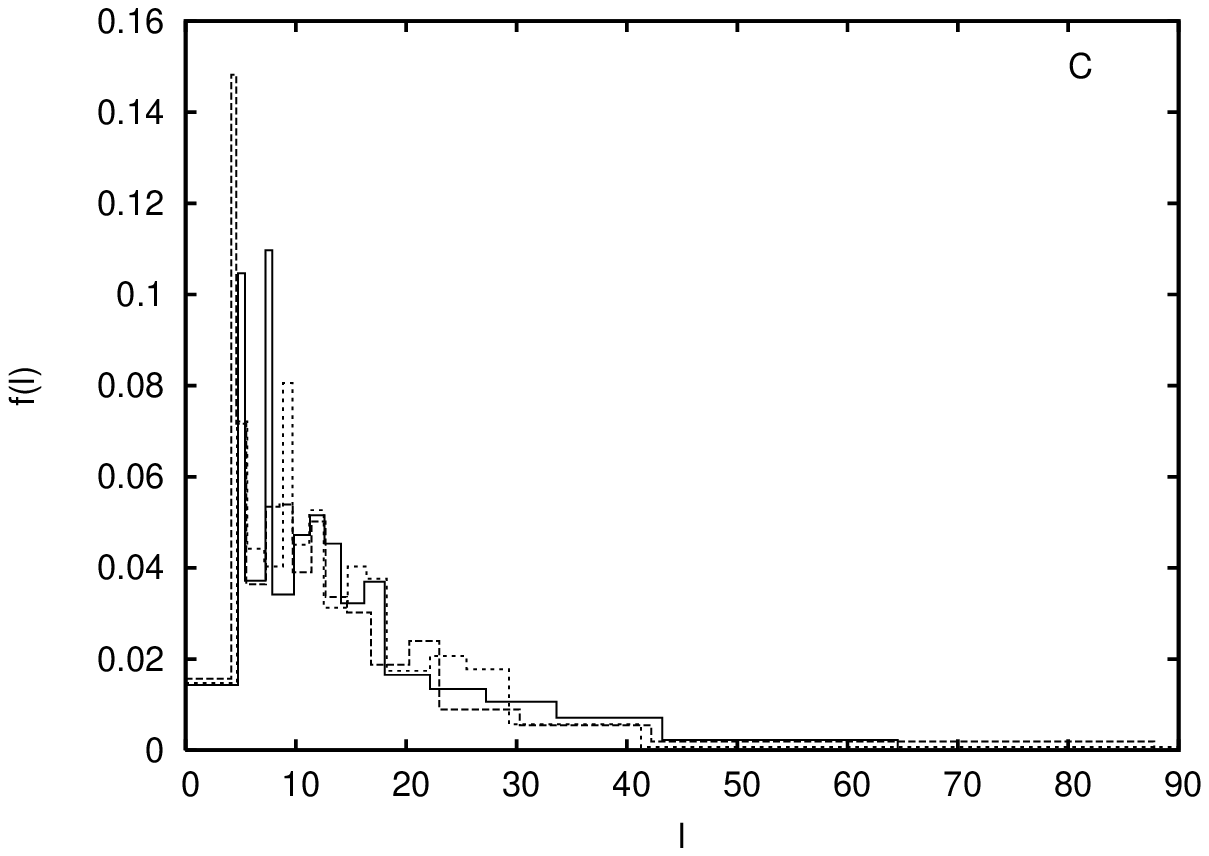}}
\resizebox{0.45\textwidth}{!}{\includegraphics*{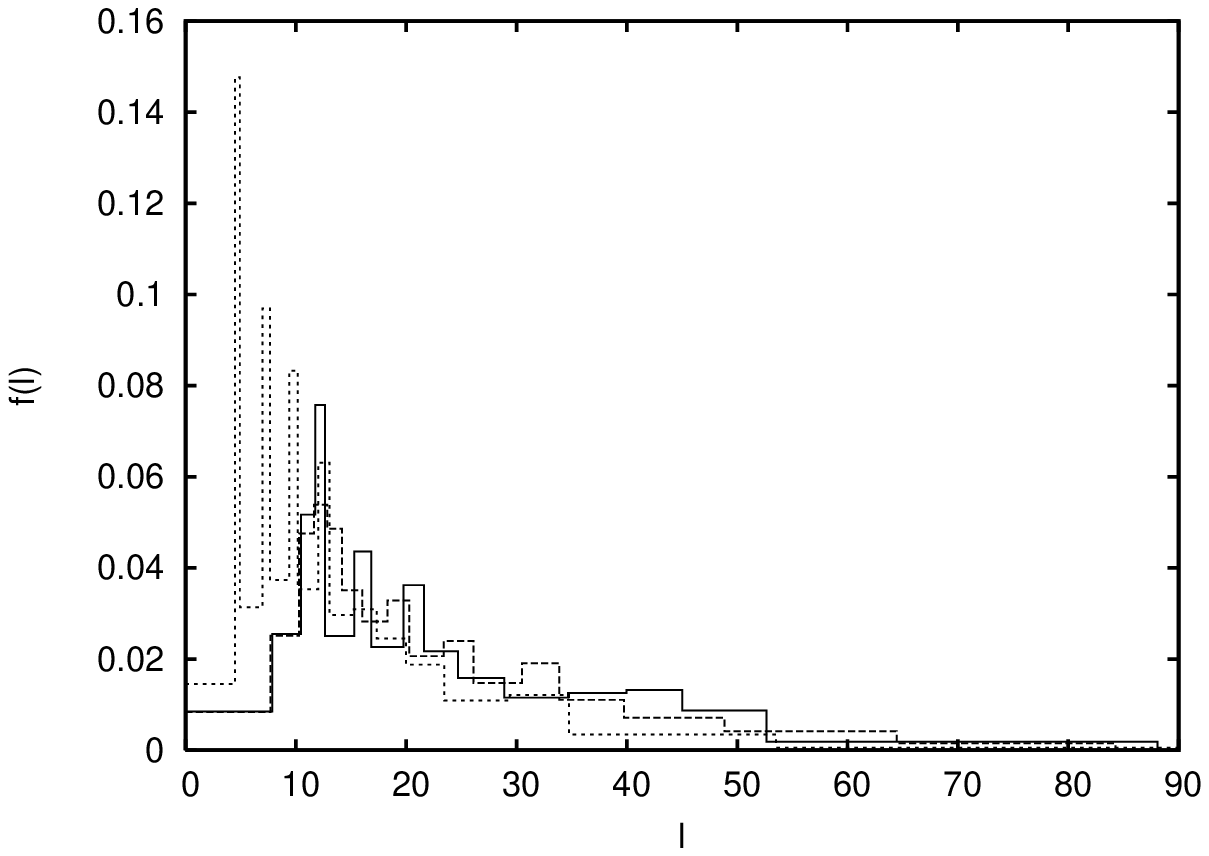}}\\
\caption{Filament length distribution histograms for the three
data sets (marked in the panels). Solid lines indicate 
parameter set 1, dashed lines set 2, and dotted lines set 3. The 
combined filament length distribution histograms for the
three data sets is shown in the bottom right panel. 
Solid line indicates data set A, dashed line
set B, and dotted line set C.
\label{fig:singhist}}
\end{figure*}

\begin{table}
\caption{The interaction parameters. \label{tab:cpar}}
\begin{center}
\begin{tabular}{|l|r|r|r|}
\hline
& \multicolumn{3}{c|}{\strut Sets}\\
\cline{2-4} \raisebox{1ex}[0pt]{Parameters}&\strut 1&2&3\\ \hline
\strut $-\log\gamma_d$ & $- 10$ & $- 10$ & $-10$\\ $- \log\gamma_f$
& 8 & 7 &
7\\ $- \log\gamma_s$ &  2 & 2 & 1\\ $- \log\gamma_o$ & 3 & 3 & 3\\
\strut $\quad V_{\max}$ & 25 & 25 & 25 \\ \hline
\end{tabular}
\end{center}
\end{table}

A simulated annealing algorithm was implemented based on
Metropolis-Hastings dynamics. The parameter for the cooling scheme
was taken as $c=0.9995$ and the initial temperature was set to $10$.
The algorithm was run for $10^7$ iterations, and the temperature
was lowered every $10^3$ steps.

The Candy model has a large number of parameters, and these should
be chosen carefully in order to get a good representation of
the filaments in the data. The segment parameters (segment lengths and
widths) have to be chosen such that the model filaments follow those in
the data. Thus, for the first two data sets, the segment parameters were
$l_{\min}=3,l_{\max}=5,w_{\min}=1,w_{\max}=2$; for the third data
set, smaller segments were considered:
$l_{\min}=2,l_{\max}=3,w_{\min}=0.95,w_{\max}=1.05$ (all distances
are given in $h^{-1}$ Mpc). The interaction regions were defined by
choosing the radius of the connecting region $r_c=0.5$ and the
rejection parameter that forbids segments to cross, $\delta=0.1$
radians. The orientation parameter, which limits the local curvature
of filaments, was chosen to be $\tau=0.5$ radians for the first two
data sets and to be $\tau=0.75$ radians for the data set C.

We experimented with a large number of interaction parameters. Here
we show the results for the three sets; they give almost equally
good results. The interaction parameters for these sets are given in
Table~\ref{tab:cpar}. The optimization method was run for each data
set. High potentials were given to undesired configurations such as
single and free segments, badly aligned pairs of segments with
respect to the parameter $\tau$, and badly placed segments with
respect to the data term.

The best filament network and the comparison between the three networks
for each set of data is shown in
Fig.~\ref{fig:ds1}. Note that we do not use the
periodicity of the data --- although numerical simulations are
mostly periodic, the real galaxy distribution is not. Thus there is no
sense in complicating the numerical procedure.

The best set of parameters for data set A was set 1. Examining
the top row of Fig.~\ref{fig:ds1} we see that the 
procedure works well. All obvious
filaments that one would draw by eye are found, and the placement
of ``secondary'' filaments in more sparsely populated regions is
also good. Note also that galaxy concentrations (``clusters'') do
not destroy the filamentary pattern; filaments usually branch in
these regions.

The difference between the sets is slight, all parameter sets
represent the network fairly well. All strong filaments coincide,
the difference is in the small and weak filaments, built on a few
points only. This is well seen in the right panel, where all three
networks are superposed. Parameter set 2, for instance, generates
spurious filaments in the sparsely populated upper central region 
of the point distribution, and set 3 produces several very short isolated
filaments. On the other hand, it also provides a perfect branching
point at $x=90,\,y=30$, which sets 1 and 2 do not find.

The middle row of Fig.~\ref{fig:ds1} illustrates the 
filament networks found for 
data set B. This data set has a richer and more uniform selection of
filaments than set A. As these sets have approximately the same
number of galaxies, individual filaments in set B are more
sparse and harder to identify. Nevertheless, the method works well,
especially for the parameter set 1, the best set; this network is
shown in the middle left panel of Fig.~\ref{fig:ds1}. There are only a few
questionable short filaments, e.g. around $x=70,\,y=120$ and
$x=10,\,y=50$. Parameter set 2 generates considerably more short
isolated filaments, which do not represent the data well, and the
filaments for parameter set 3 tend to deviate in wrong
directions.

Data set C has the richest set of filaments. These are shorter
and not as pronounced as the filaments in the first two data sets
--- this is the way the large scale structure develops in the
universe. The early structure that set C describes evolves by
concentrating into larger and larger clusters and filaments;
small-scale structure becomes weaker and disappears gradually. In order
to apply the Candy model, we had to generate about twice as many
galaxies for this set as for the other two.  As shown in
the bottom row of Fig.~\ref{fig:ds1}, our procedure delineates 
the filamentary network
satisfactorily here, too, although probably the segments should
have been even smaller. As seen in the bottom left panel for the best
parameter set (set 1 in this case, too), segments sometimes jump from
an obvious filament to another (e.g., at $x=47,\,y=20$); there is
also a tendency to form short filaments for a collection of a few
points, as at $x=7,\,y=60$ and $x=75,\,y=107$.

Parameter set 2 is in this case about as good as set 1; it
misses a few obvious filaments, however (e.g., at
$x=124,\,y=50$), and has difficulties in resolving interaction
regions (knots in the network), see the region at $x=70,\,y=110$.
This region has been equally difficult to model for all three
parameter sets. And, finally, parameter set 3 gives the worst
filament placement among the three.

\subsection{Length distribution}

As the Candy model is able to reconstruct the filamentary network,
given a point process (galaxy map), the collection of its
parameters can be considered as a description of the network. When
determined from the data by a likelihood procedure, they can serve
as statistics of the network. But there are simple statistics we
can already study; the simplest one is the probability
distribution of the lengths of individual filaments (sets of
connected segments). A similar problem, that of the length of the
largest filament, has been addressed recently, using a pixel-based
method to define filaments and finding the pixel size where the
filaments are still statistically significant \citep{bhavsar03}. As
filaments delineate voids, the distribution of the lengths of
filaments is also connected with the distribution of void sizes.
This subject has a long history; see, e.g., \citet{martsaar02} and
an interesting recent theoretical paper by \citet{weyg03}.

Comparison of the length histograms in Fig.~\ref{fig:singhist}
reveals a series of peaks, several distinct characteristic lengths
in the filamentary network\footnote{The histograms shown have bins
of varying width since they are of equal area.
Although not common, it is a good, non-parametrical way to represent
probability densities.}. These peaks are especially prominent for
data sets A and C, and smeared out for set B. Also, the
locations of the peaks do not depend much on the specific parameter
set, the peaks more or less coincide. And although the sample sizes
are a bit too small for the first two data sets to draw firm
conclusions, inspection of the probability distributions
shows that the peaks are real.

Another feature of the distributions is their pronounced tails -- the
lengths of filaments reach about 90, almost the full size of the
box. Inspection of Fig.~\ref{fig:ds1} 
shows that long filaments are those that pass through
the branching points and are really collections of several
filaments. So, in the future we have to find a recipe for locating
the branching points and breaking the filaments; otherwise we shall
lose connection with the void distribution. For the histograms in
Fig.~\ref{fig:singhist} this means that there would be additional
contributions to the 20-40 length range, which are presently
missing.

As the locations of the peaks are almost independent of the parameter
set used, we combined the length data for the three parameter sets.
These distributions for the three data sets are compared
in the bottom right panel of Fig.~\ref{fig:singhist}. 
Thus these peaks are significant,
revealing discrete scales in the data. We also see that the overall
length distribution is shifted to the shorter sizes for set C,
compared with A and B.

\section{Other approaches}

\begin{figure*}
\centering
\resizebox{!}{0.32\textwidth}{\includegraphics*{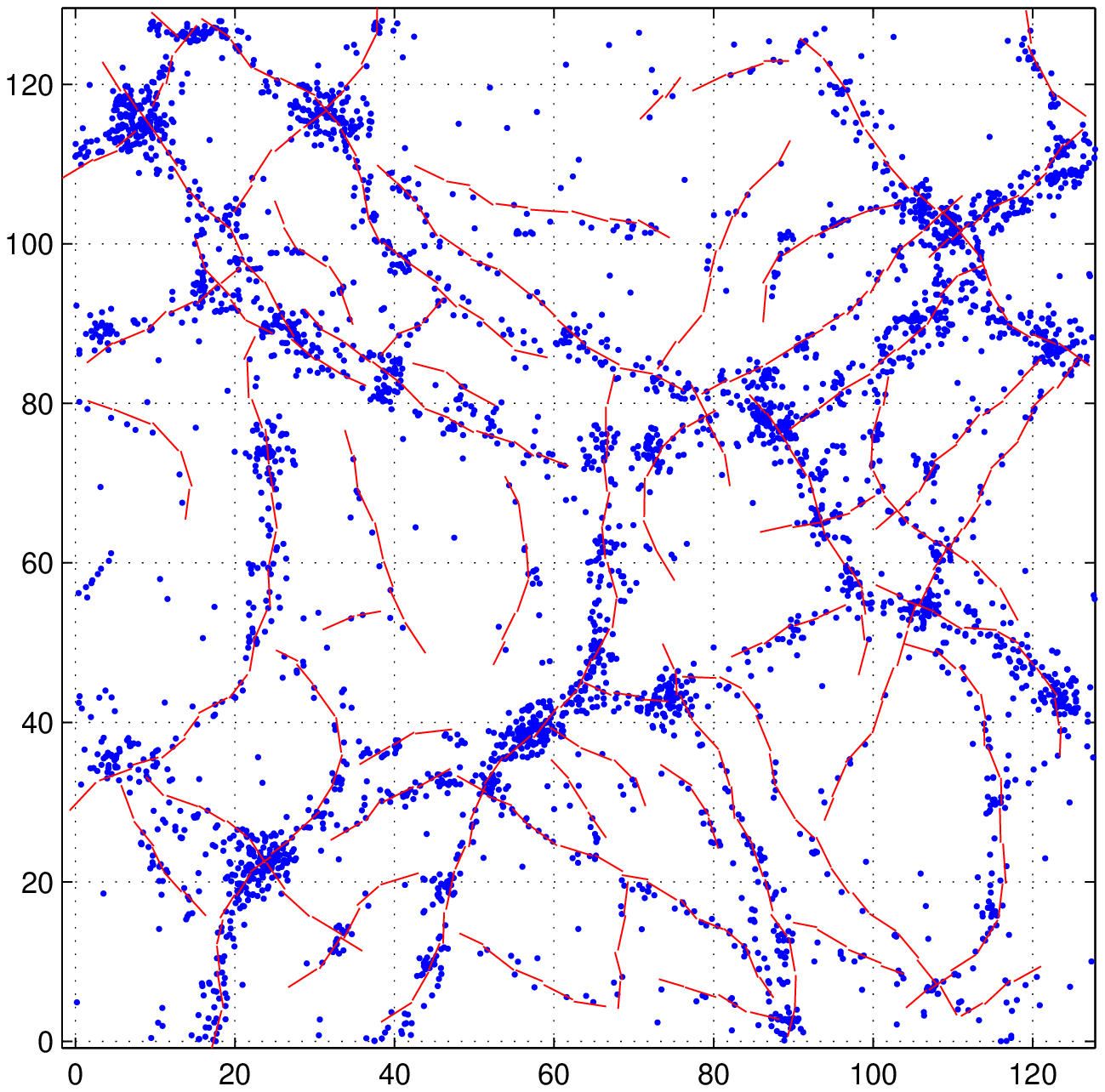}}
\resizebox{!}{0.32\textwidth}{\includegraphics*{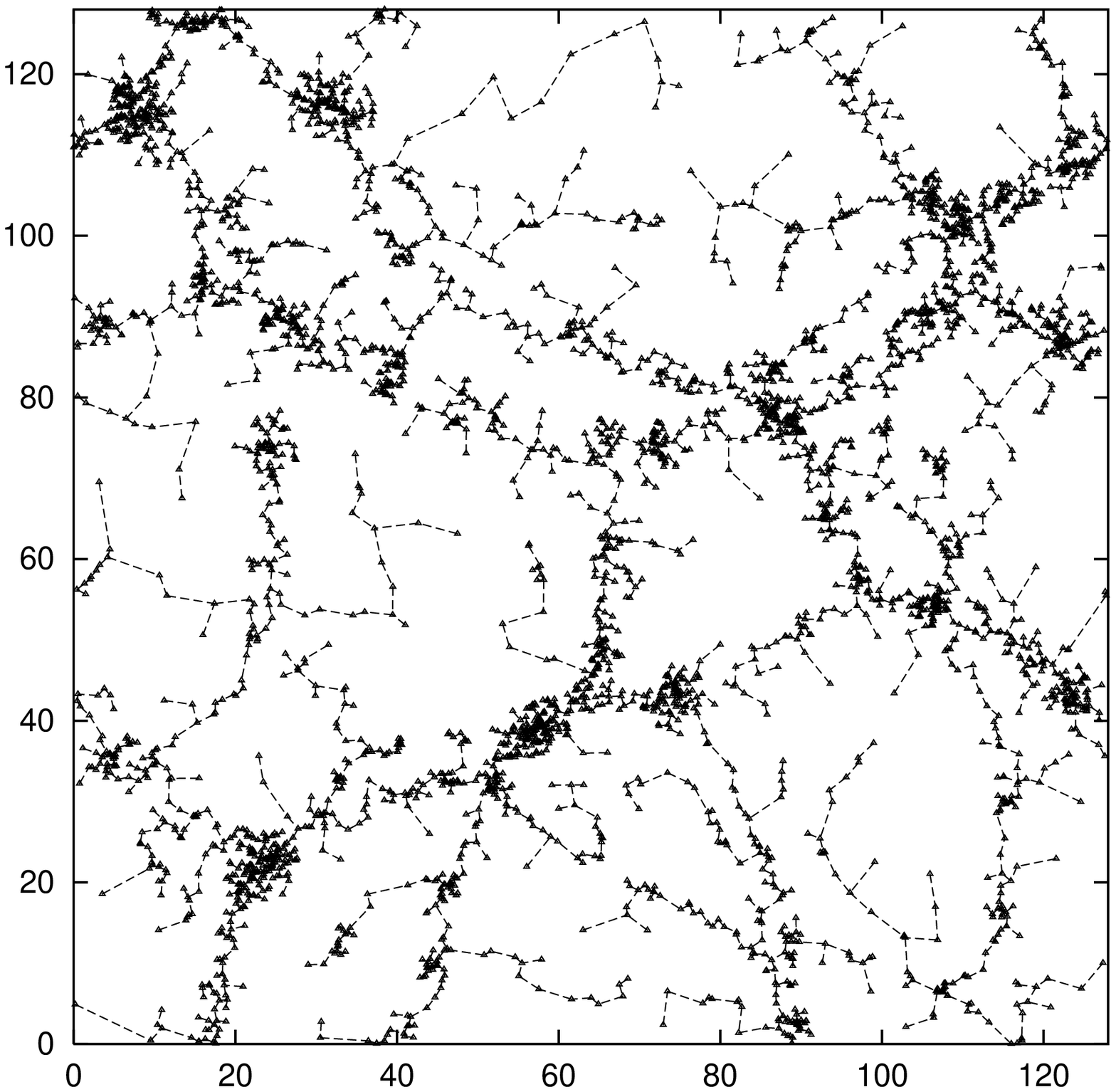}}\\
\caption{A Candy model (left panel) and the minimal spanning tree
(right panel) for the same set of data (set A)). \label{fig:mst}}
\end{figure*}

There exist only a few methods to describe the observed filamentary
network of galaxies. The best known approach is that of minimal
spanning trees (MST) \citep{barrow85}. The minimal spanning tree
connects all data points, and its length distribution function
describes mainly the nearest-neighbour distance distributions, not
the large-scale network we see. The MST also has to connect all
points in the clusters, while the Candy model can be tuned to ignore
them (as we have seen, clusters usually become branching points of
the filament network in the Candy model). The differences between
the MST and the Candy model can be well seen in Fig.~\ref{fig:mst}.
Nevertheless, the MST has been used extensively to describe the
filamentary network; as the Candy model looks much better 
and is well suited for incomplete samples, it should
lead to a better description of the cosmic filaments.

Using a technique based on multiscale geometric analysis
\citet{donoho} have showed how filaments can be
detected when they are embedded in a uniformly distributed
background of points. This algorithm is specially aimed at 
finding hidden filamentary patterns in images or nearly Poissonian
point processes. Our approach is different and is better suited for
finding many filaments in clustered point processes, because the
Candy model produces smoother maps and is able to better combine
both local and large scale characteristics of the galaxy
distribution.

Another, more recent approach to describing filaments \citep{bhavsar03}
proceeds by binning the map (calculating a density field), and using
Minkowski functionals of the isodensity contours to estimate the
filamentarity of the objects. While this approach will classify all
objects, it has two free parameters, the smoothing length (size of
the density bin), and the isodensity level.  True, in some respects
our approach is similar to that,
 as the segments of the
Candy model have a finite width (we are also estimating a density
field). But our density estimator is anisotropic and adaptive, in
principle, and we trace only filamentary structures.

A third approach that is also based on a density field, is to
determine the saddle points and to build a network of field lines
(directed along the gradient of the field), connecting saddle
points with local maxima -- the skeleton \citep{novikov03}. This
approach could reconstruct well the cellular network of filaments
(so far it has been applied to studying the pixelized cosmic
microwave background data by \citealt{eriksen04}), but it will also
depend on the density estimation procedure. And, as the authors
say, this approach is computationally complex.

\section{Conclusion and perspectives}

The parameter values for our method were chosen by trial
and error. Under these circumstances, parameter estimation using
Monte Carlo maximum likelihood methods may be considered
\citep{Geye99,LiesStoi03}. These parameters could then be considered
as statistics describing the filamentary network. They will
certainly be much better suited for this task than the moments of
the density distribution in real or in Fourier space which are
commonly used in cosmology.

The data term is a very simple test. Much more sophisticated
techniques such as testing the alignment of the points covered by a
segment, or statistical tests  such as the complete randomness
test need investigation.

To our knowledge there is no proof for the existence of an optimal
cooling scheme when using Metropolis-Hastings dynamics for
simulating point processes in a simulated annealing algorithm.
There is such a proof for the spatial birth-and-death process, but
in practice the authors sample the model using a fixed cold
temperature \citep{Lies94}. The choice we opted for, a slow
polynomial decreasing scheme, does not guarantee that the global
optimum is reached.

But overall, as we have seen, the results are good, the Candy
model can be tuned to trace the filamentary network well. And it
can be naturally generalized to describe the real 3-D filamentary
networks of galaxy maps; see \citet{gregory04}. As we already said
above, it can also be considered as a tool for providing statistics
of filamentary networks. These are the future directions of our
work.

\section{Acknowledgements}
We thank the referee Kevin Pimbblet 
for interesting comments that improved the paper.
Enn Saar and Radu Stoica  want to thank the
Observatori Astron\`omic de la Universitat de Val\`encia where part
of this work was done for its hospitality. 
This work has been supported by Valencia
University through a visiting professorship for Enn Saar, by the
Spanish MCyT project AYA2003-08739-C02-01 (including FEDER), by the
Generalitat Valenciana project GRUPOS03/170, and by the Estonian
Science Foundation grant 4695. We thank Rien van de Weygaert for his
code to calculate the MST. The work of Jorge Mateu and Radu Stoica
was carried out, respectively, under the grants BFM2001-3286 of the
Spanish MCyT and SB2001-0130 from MECD. 

\bibliographystyle{aa}
\bibliography{candyaa}

\end{document}